\documentclass[9pt,twocolumn,twoside]{osajnl}

\journal{pr} 

\setboolean{shortarticle}{true}

\title{Multiphoton ionization of standard optical fibers}

\author[1,2,\dag,*]{M. Ferraro}
\author[1,3,\dag]{F. Mangini}
\author[1]{Y. Sun}
\author[1]{M. Zitelli}
\author[3]{A. Niang}
\author[2]{M.C. Crocco}
\author[2]{V. Formoso}
\author[2]{R.G. Agostino}
\author[2]{R. Barberi}
\author[2]{A. De Luca}
\author[4]{A. Tonello}
\author[4]{V. Couderc}
\author[5,6]{S.A. Babin}
\author[1,6]{S. Wabnitz}

\affil[1]{Department of Information Engineering, Electronics and Telecommunications, Sapienza University of Rome, Via Eudossiana 18, 00184 Rome, Italy}
\affil[2]{Physics Department and STAR infrastructure, University of Calabria, I-87036 Arcavacata di Rende, CS, Italy}
\affil[3]{Department of Information Engineering, University of Brescia, Via Branze 38, 25123 Brescia, Italy}
\affil[4]{Universit\'e de Limoges, XLIM, UMR CNRS 7252, 123 Avenue A. Thomas, 87060 Limoges, France}
\affil[5]{Institute of Automation and Electrometry, SB RAS, Novosibirsk 630090, Russia}
\affil[6]{Novosibirsk State University, Pirogova 1, Novosibirsk 630090, Russia}

\affil[$\dag$]{These authors have contributed equally}
\affil[*]{Corresponding author: mario.ferraro@uniroma1.it}

\begin{abstract}
Atoms ionization by the simultaneous absorption of multiple photons has found applications in fiber optics, where it leads to unique nonlinear phenomena. To date, studies of the ionization regime have been limited to gas-filled hollow-core fibers. Here, we investigate multiphoton ionization of standard optical fibers, where intense laser pulses ionize the atoms constituting the fiber structure itself, instead of that of the filling gas. We characterize material modifications produced by optical breakdown. Their formation affects laser beam dynamics over hours long temporal scales. The damage features are studied by means of optical microscopy and X-ray microtomography. In the framework of glass photonics, our results pave the way for a novel glass waveguide micromachining technique.
\end{abstract}

\setboolean{displaycopyright}{true}

\begin{document}
\maketitle
The ionization of atoms has been studied for centuries, and it has been experimentally demonstrated in atomic gases. Analogously, ionization can be achieved in solids, and specifically dielectrics. The free electrons which result from ionization generate a plasma, whose ignition produces modifications of the material, i.e., structural damages ~\cite{davis1996writing}. Properly engineering these modifications allows for structuring glasses, giving rise to the field of micromachining ~\cite{hanada2004development}. Specifically, modifications made on the glass surface can result in the removal of the material, which is referred to as optical ablation ~\cite{stuart1996optical, chimier2011damage}. 

Since ionization requires the extraction of electrons from the material, photons must have energies higher than the material bandgap, which in glasses is of the order of a few electronvolts. Therefore, at first ultraviolet (UV) lasers were used for material ionization \cite{karlitschek1995photodegradation}. However, the latter can also be obtained when employing lasers with infrared (IR) wavelengths, by exploiting nonlinear effects based on multiphoton absorption (MPA) ~\cite{lenzner1998femtosecond}. Although this requires laser intensities as high as a few TW/cm$^2$, multiphoton ionization (MPI) provides many advantages. Notably, MPI permits for reducing the energy which is needed for achieving structure modifications, thus reducing their size, and improving the precision of laser-based micromachining and ablation. 

Nevertheless, MPI has also some drawbacks. Notably, when dealing with high peak power IR lasers, due to the presence of thermal effects,
one may risk to not be able of controlling the features of the induced damages, e.g., their size or rate of formation. In order to overcome this issue, the use of femtosecond laser pulses has been proposed. Ultrafast optics allows, in fact, to avoid detrimental thermal effects \cite{perry1999ultrashort}.
Although the MPI regime has been widely studied in bulk glasses ~\cite{von1996breakdown, li1999ultrafast}, studies of the ionization regime in optical fibers have been limited to the case of photonic bandgap guiding structures ~\cite{holzer2011femtosecond,saleh2011theory}. In the latter, a plasma is generated thanks to the ionization of a gas that fills the hollow fiber.

As a matter of fact, only few works dealing with standard optical fibers in the MPI regime have been reported in the literature. It is well-known that MPI enables the fabrication of fiber Bragg gratings and, when combined with plasma filamentation, may permit the manufacturing of 3D optical memories \cite{cho1998observation}. Furthermore, it has been recently shown that the cylindrical geometry of optical fibers can be exploited, in order to provide non-straight shapes to such plasma filaments \cite{mangini2020spiral, mangini2022helical}. Nevertheless, to our knowledge, a full characterization of the effects of laser-induced damages on fiber properties and a detailed study of the damage features have not been reported so far. 

In this work, we fill this lack by investigating the MPI regime in standard multimode optical fibers (MMFs). The large core area of the latter make them more suitable than singlemode fibers for exploring extreme nonlinear effects \cite{picozzi2015nonlinear,krupa2019multimode}. Specifically, we study the effects of MPI-induced damages on the main guiding properties of MMFs. Remarkably, we found that the process of multimode supercontinuum generation (SCG) undergoes a complex evolution on a time scale of several hours. Such an evolution is irreversible, and it comes along with a drop of the net optical transmission from the MMF. Numerical simulations provide useful insight of the physical mechanisms underlying SCG. Since standard optical fibers are mainly used around 1.5 $\mu$m for telecom applications, we firstly investigated SCG in this wavelength range. However, in order to emphasize the role of absorption, which leads to the material damages, we also carried out experiments at 1 $\mu$m, the wavelength of operation of Yb-based fiber lasers, whose energy up-scaling and mode-locking remains an open challenge \cite{krupa2019multimode,wright2017spatiotemporal}. Finally, we analysed post-mortem samples of MMFs by means of optical microscopy and computed X-ray micro-tomography ($\mu$-CT). The latter has been recently proposed as a tool for imaging optical fibers \cite{sandoghchi2014x,levine2019multi,levine2021x}. In this framework we demonstrate, we believe for the first time, that $\mu$-CT can be used for studying the 
MPI-induced modifications of MMFs.

In our experiments, we used graded-index (GRIN) MMFs, whose linear refractive index $n$ has a parabolic profile inside the core:
\begin{equation}
    n = n_0 [1 - g (x^2+y^2)],
    \label{grin}
\end{equation}
where x and y are transverse coordinates, orthogonal to the fiber axis.
GRIN fibers are, in fact, best suited for studying the MPI regime, since they require smaller input energies than step-index MMFs for triggering MPA effects~\cite{ferraro2021femtosecond}. The fiber core radius and the nominal grading index factor values provided by the manufacturer are $r_c$=25 $\mu$m and $g
= 3.29 \cdot 10^{-5} (\mu m)^{-2}$, respectively.
Our source consists of an ultra-short laser system, involving a hybrid optical parametric amplifier (OPA) of white-light continuum (Lightconversion ORPHEUS-F), pumped by a femtosecond Yb-based laser (Lightconversion PHAROS-SP-HP), generating pulses at 100 kHz repetition rate and $\lambda=1.03$ $\mu$m, with Gaussian beam shape ($M^2$=1.3). The OPA allows for converting the laser wavelength to $\lambda=1.55$ $\mu$m
. The pulse shape was measured by using an autocorrelator (APE PulseCheck type 2), resulting in a sech temporal shape with pulse width of 67 fs for the OPA, and 174 fs for the main source.

Laser pulses were injected into the MMF by a convex lens, so that at the fiber input the beam $1/e^2$ of peak intensity is approximately 15 $\mu$m and 30 $\mu$m at $\lambda = 1.03$ $\mu$m and $\lambda = 1.55$ $\mu$m, respectively.
During material damaging, the input tip of the fiber is imaged by a digital microscope (Dinolite-AM3113T), whereas 
at the fiber output, a micro-lens collimates the out-coming beam, which is sent through a cascade of beamsplitters to a broadband thermopile power meter (Gentec XLP12-3S-VP-INT-D0) and a real-time multiple octave spectrum analyzer (Fastlite Mozza) with an operating wavelength range of 1.0-3.0 ~$\mu$m.

\begin{figure}[ht!]
\centering\includegraphics[width=8.7cm]{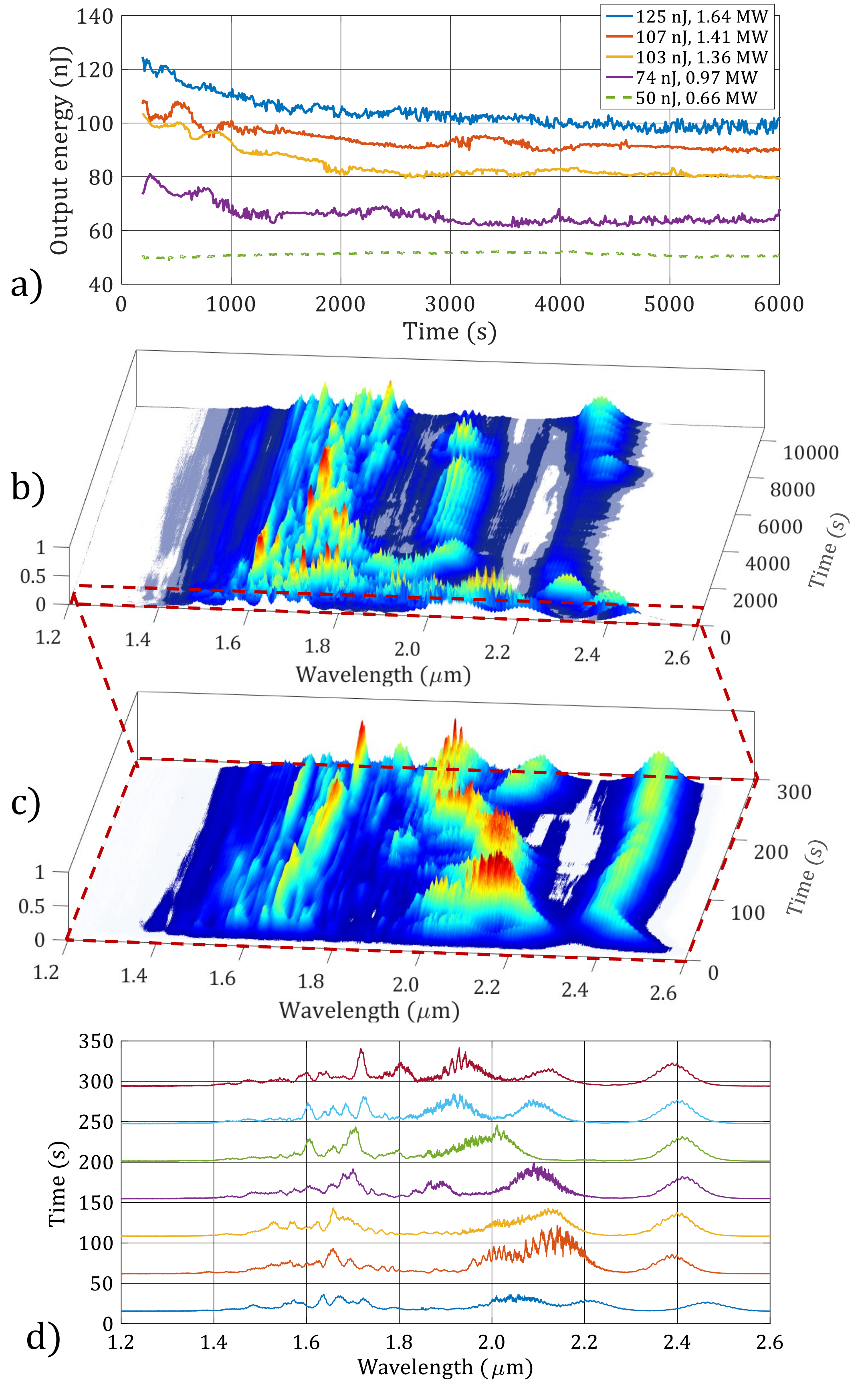}
\caption{(a) Time evolution of the output beam power from a 30 cm long GRIN fiber span, for different values of the input peak power. The legend shows the values of input energy and peak power, respectively. (b) Slow temporal evolution of the output spectrum, for fixed input pulse peak power P = 1.41 MW. (c) Zoom-in of (b), for the first 285 s. (d) Same spectra of (c) at 7 temporal instants of time.
}
\label{dance}
\end{figure}

When working with high peak power femtosecond pulses in the anomalous dispersion regime of GRIN MMFs, e.g., at $\lambda=1.55$ $\mu$m, multiple fundamental solitons are generated via the fission of the input multisoliton pulse ~\cite{zitelli2020high}. These solitons are peculiar solutions of the coupled nonlinear Schr\"odinger equations, which propagate undistorted thanks to compensation of fiber dispersion by nonlinearity. As well known from the singlemode case, fundamental solitons have a sech shape both in the temporal and in the spectral domains. Moreover, it has been shown that, due to their intrinsic multimode nature, these solitons have unique features, e.g., their pulse width (which is of the order of 100 fs) barely depends on their energy~\cite{zitelli2020high}. Owing to the Raman effect, the solitons undergo a continuous red-shift, which quadratically increases with their energy, also known as Raman-induced soliton self-frequency shift (SSFS) ~\cite{santhanam2003raman, ahsan2018graded}.
However, as the SSFS-induced wavelength shift grows larger, silica absorption becomes more relevant. As a result, the soliton wavelength is clamped to a maximum value, which depends on the fiber length. Specifically, for fiber spans of up to a few tens of meters, linear fiber loss limits the maximum soliton wavelength to $\lambda_M \simeq 2.2$ $\mu$m.
On the other hand, when employing centimeter scale fiber samples, the soliton wavelength may grow larger beyond 2.1 $\mu$m, if enough input pulse energy is provided. Nevertheless, $\lambda_M$ cannot be increased at will. Above a certain input pulse energy threshold, MPA becomes non-negligible, thus clamping the maximum output soliton energy and, consequently, its red shift~\cite{zitelli2020high}. 
Energy dissipated by MPA turns out to excite material defects, some of which re-emit light in the form of visible (VIS) luminescence \cite{hansson2020nonlinear}. In this regime, the defects are able to absorb all of the dissipated energy, so that no permanent damages are produced to the fiber. This becomes clear by monitoring the evolution of output power vs. time. Whenever the input power is sufficiently low, the output power remains constant in time (see dashed curve in Fig.\ref{dance}a, obtained at $P = P_{thr} = 0.97$ MW). On the other hand, as soon as the input power overcomes a certain threshold, or $P > P_{thr}$, the power transmission of the fiber slowly drops in time. The output energy is reduced to about 80\% of its initial value after a few hours (see the curves in Fig.\ref{dance}a at $P>4.48$ MW). Interestingly, the transmission drop time scale appears to be largely independent on input power: all of the solid curves in Fig.\ref{dance}a exhibit the same power damping timescale. We underline that the power values used in our experiments are smaller than the power value for critical self-focusing $P_{csf}(\lambda = 1.55$ $\mu$m) = 9.6 MW, which is calculated as $P_{csf} = \lambda^2/2\pi n_0 n_2$. 


The irreversible reduction of the output power from the fiber is the hallmark of damage generation: these are progressively formed, and are responsible for a variation of the local refractive index. As it can be seen in Fig.\ref{dance}a, the power transmission curve shows several fluctuations, before reaching a steady value after about 2 hours. 
As a matter of fact, SCG from the GRIN fiber exhibits a slow evolution, which is characterized by two different time scales. These cannot be distinguished by the thermal power meter, because of its slow response. However, one can easily monitor variations of the SC spectra at the fiber output. For the sake of simplicity, we only report in Fig.\ref{dance}b and c the temporal dependence of the output spectrum, which corresponds to the red curve in Fig.\ref{dance}a. As one can see in Fig.\ref{dance} c, SCG exhibits a slow and continuous reshaping: 
two distinct spectral lobes, corresponding to different Raman solitons that are initially generated around 2000 nm and 2200 nm, respectively, merge into a single lobe after 50 s, only to split again after about 200 s. 
Eventually, Fig.\ref{dance} a,b show that, over a time scale of hours, the output power remains stable; correspondingly, an equilibrium spectral distribution is reached, with $\lambda_M \simeq 2.2$ $\mu$m. 
Whereas, for times shorter than 100 s, $\lambda_M$ decreases quickly from an initial value of 2.5 $\mu$m down to about 2.4 $\mu$m (see Fig.\ref{dance}c). In order to highlight the SCG evolution on the shortest time scale, in Fig.\ref{dance}d we report a 2D plot of the output spectrum at specific (equally spaced) instants of time.

\begin{figure}[ht!]
\centering\includegraphics[width=8.7cm]{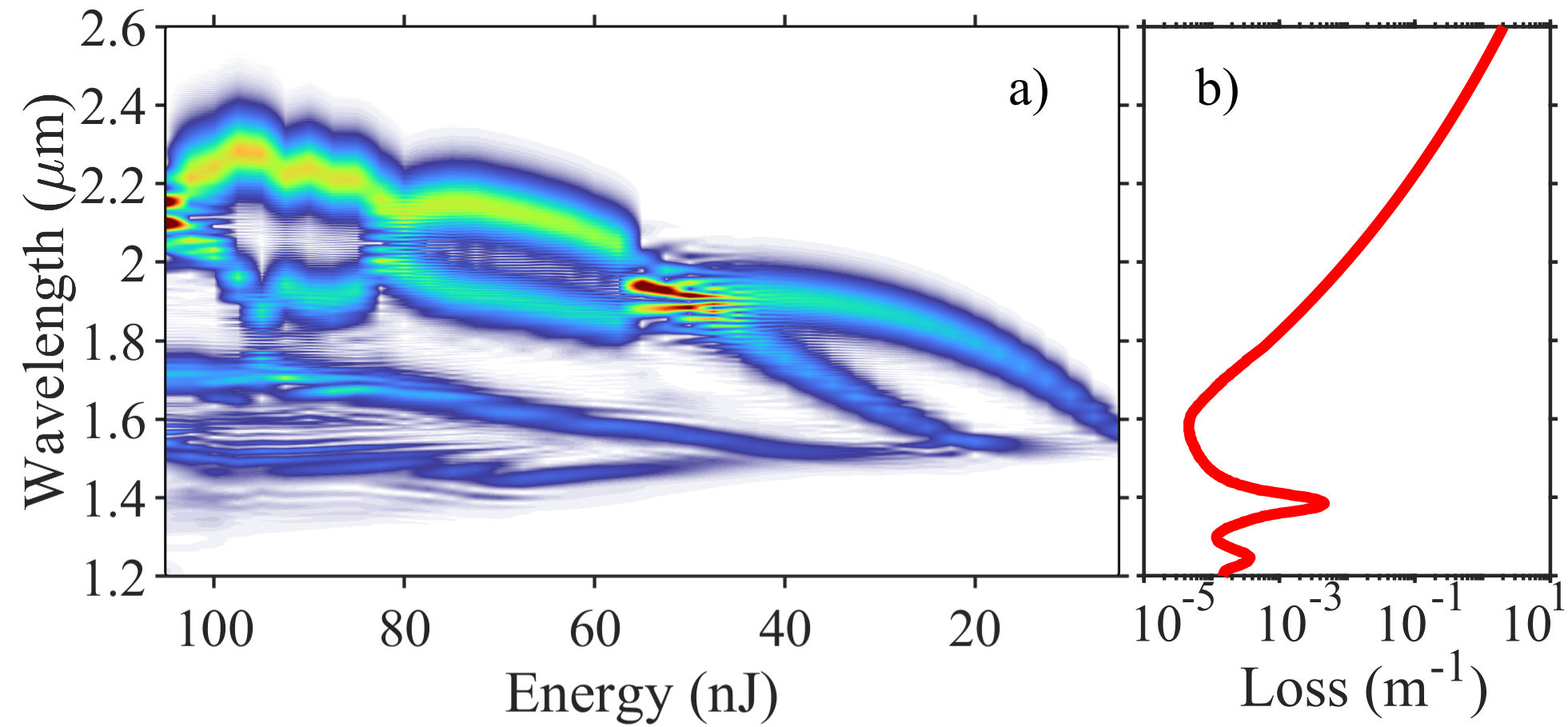}
\caption{(a) Numerical simulation of output spectrum changes vs. input pulse energy. (b) Wavelength dependence of the linear loss profile in simulations. All parameters are the same as in experiments of Fig.\ref{dance}. 
}
\label{simulations}
\end{figure}

The occurrence of a power transmission drop and its associated stable output spectral distribution are due to a quenching of the output beam energy, owing to an increase of either fiber absorption or light scattering by the generated damages \cite{reupert2019side,rawson1972measurement}.
Indeed, our experimental observations can be qualitatively reproduced by numerical simulations based on the generalized multimode nonlinear Schr\"odinger equation: model details can be found in Ref.\cite{wright2017multimode}. Here, we include the presence of wavelength dependent linear losses: the fiber loss coefficient increases by four orders of magnitude when $\lambda$ grows from 1.55 $\mu$m up to 2.1 $\mu$m (see Fig.\ref{simulations}b). On the other hand, we did not include in simulations any nonlinear term for describing the presence of MPI, since its contribution to the spatiotemporal beam dynamics acts over a time-scale which is much shorter than that associated with damages formation \cite{hamam2018numerical}. 
To the contrary, we could mimic the process of damage generation by just decreasing the energy of the input laser pulses. As it turns out, this permits for qualitatively retrieving the experimentally observed changes of the output spectra. As can be seen, simulations reported in Fig.\ref{simulations}a show that multiple spectral lobes, associated with the different Raman solitons that are generated at the fiber output, separate, merge, and then separate again, as the input pulse energy grows smaller. This is similar to the observations of Fig.\ref{dance}b, when we consider that fiber losses progressively grow larger in time. A detailed study of the process of soliton spectral merging, whose discussion is beyond the scope of the present work, reveals that it can be explained by considering the multimode nature of Raman soliton generation in a GRIN fiber. 

\begin{figure}[!ht]
\centering\includegraphics[width=8.7cm]{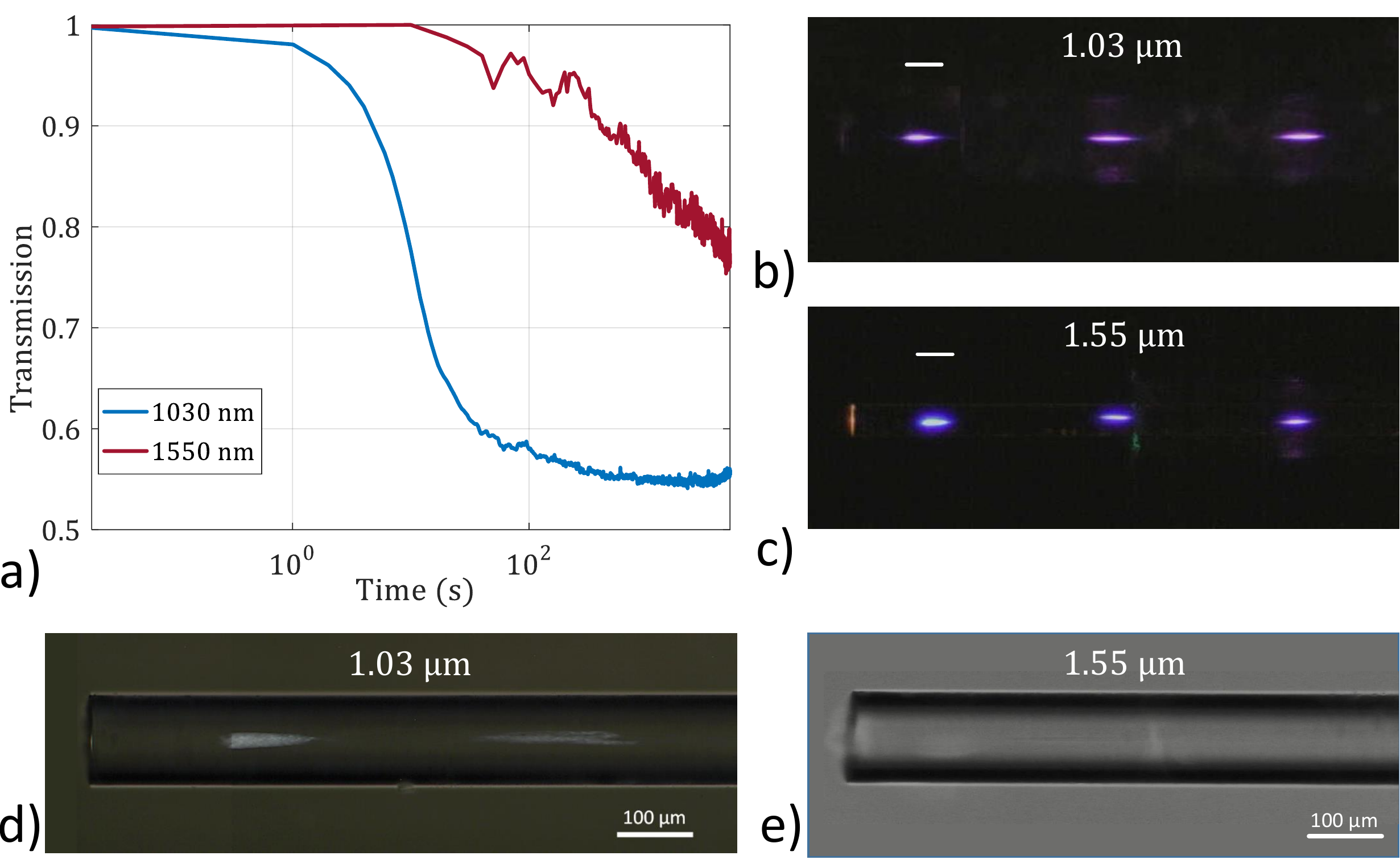}
\caption{Comparison between MPI regime established at 1.03 $\mu$m and 1.55 $\mu$m of wavelength: (a) transmission, (b,c) luminescence, (d,e) damages imaged by optical microscopy. The white bars in (b-e) correspond to 100 $\mu$m. The input energy and the peak power are E = 323 nJ and P = 1.64 MW at 1.03 $\mu$m and E = 125 nJ and P = 1.64 MW at 1.55 $\mu$m, respectively. The images in d,e) have been obtained after 1.5 h of exposition to the laser.
}
\label{comparison}
\end{figure}

Although we have shown that MPI-associated defects have a dramatic effect on SCG pumped at 1.55 $\mu$m, at this wavelength linear loss has a minimum (see Fig.\ref{simulations}b), hence MPI is relativey weak. Therefore, in order to maximize the generation of laser-induced damages, we moved the pump pulse wavelength to 1.03 $\mu$m, so that its propagation occurs in the normal dispersion regime. 
Remarkably, although changing the dispersion regime at the pump wavelength has profound effects on SCG, the fiber transmission drop at 1.03 $\mu$m turns out to be similar (and much more rapid) than at 1.55 $\mu$m (cfr. Fig.\ref{comparison}a). This indicates that most of pump power is lost at the beginning of the fiber, i.e., over the first few millimeters. As a matter of fact, by optical microscopy we could confirm that, at longer fiber lengths no damages were generated. Furthermore, we verified that by cleaving the fiber input face by about 5 millimeters, we could fully recover the original level of fiber transmission. 

Note that the millimeter scale is much shorter than the nonlinear length associated with other nonlinear effects besides MPI, such as SSFS and SCG. This can be appreciated by looking at Fig.\ref{comparison}b and c, where we compare the MPA-induced luminescence of the fiber input tip, when using input fs pulses at either 1.03 $\mu$m or 1.55 $\mu$m pump wavelengths, respectively. As it can be seen, the luminescence traces looks rather similar, and no effects of VIS light emission that is associated with SCG can be detected. Only slight differences in the luminescence color are appreciable: these can be ascribed to the different nature of the laser-induced defects \cite{mangini2020multiphoton}. Luminescence appears at equally spaced discrete points, owing to spatial self-imaging, a typical property of GRIN fibers \cite{karlsson1992dynamics}: because of the refractive index grading, the propagating beam periodically reproduces its spatial profile. 
As a side note, it is worth to mention that luminescence is a priceless tool for micromachining, since it allows for a significant improvement of the laser beam alignment. This has found applications to the inscription of fiber Bragg gratings \cite{hnatovsky2017nonlinear}, for determining the optical fiber cut-off frequency \cite{mangini2021experimental}, as well as for tracking the propagation path of skew-rays \cite{mangini2020spiral}.


The difference between defect generation by laser pulses propagating in opposite chromatic dispersion regimes can be better appreciated by imaging post-mortem samples. In Fig.\ref{comparison}d and e, we show images captured by an optical microscope (Zeiss Axiolab 5) in a quasi-cross polarization configuration. As it can be seen, although exactly the same input conditions (peak power and beam waist) were used, the fiber is notably more damaged when the input wavelength is 1.03 $\mu$m (Fig.\ref{comparison}d). Whereas the fiber appears to be barely affected by pump pulses at 1.55 $\mu$m (Fig.\ref{comparison}e). The microscope image in Fig.\ref{comparison}d allows for clearly appreciating the shape of the damages, which is remarkably similar to that of MPI-ignited plasma at the self-focusing point in bulk glasses \cite{schmitz2012full, sudrie2002femtosecond}. It is worth to point out that, at variance from the latter, where damages are only induced at the self-focusing point, when the guiding properties of MMFs are exploited (specifically, the spatial self-imaging effect), one is able to simultaneously modify the refractive index of the fiber core along an array of several (and equally spaced) points. 


Optical microscopy techniques, such as shearing, phase-contrast, and Fourier, as well as Raman spectroscopy are largely used to detect the presence of damages in glasses \cite{chan2001structural,osellame2012femtosecond}. However, although quantitative information can be extracted about the absolute value of the refractive index local variation, the sign of this variation remains unknown. As a matter of fact, power collected by the microscope scales quadratically with the laser-induced refractive index change \cite{reupert2020angular}. 

In this context, we propose to exploit the material properties at X-ray frequencies. Here, $\mu$-CT becomes an efficient tool for gathering additional information about the refractive index variation. It is worth to highlight that conventionally $\mu$-CT finds applications in different fields, e.g., material science \cite{conte2019analysis}, geophysics and geology \cite{cnudde2013high}, soil science \cite{taina2008application} and archaeology and cultural heritage \cite{stabile2021computational, lopez2021architectural}. To the best of our knowledge, this is the first time that X-ray micro-tomography is used for studying laser-induced damages of optical fibers. 

\begin{figure}[ht!]
\centering\includegraphics[width=8.6cm]{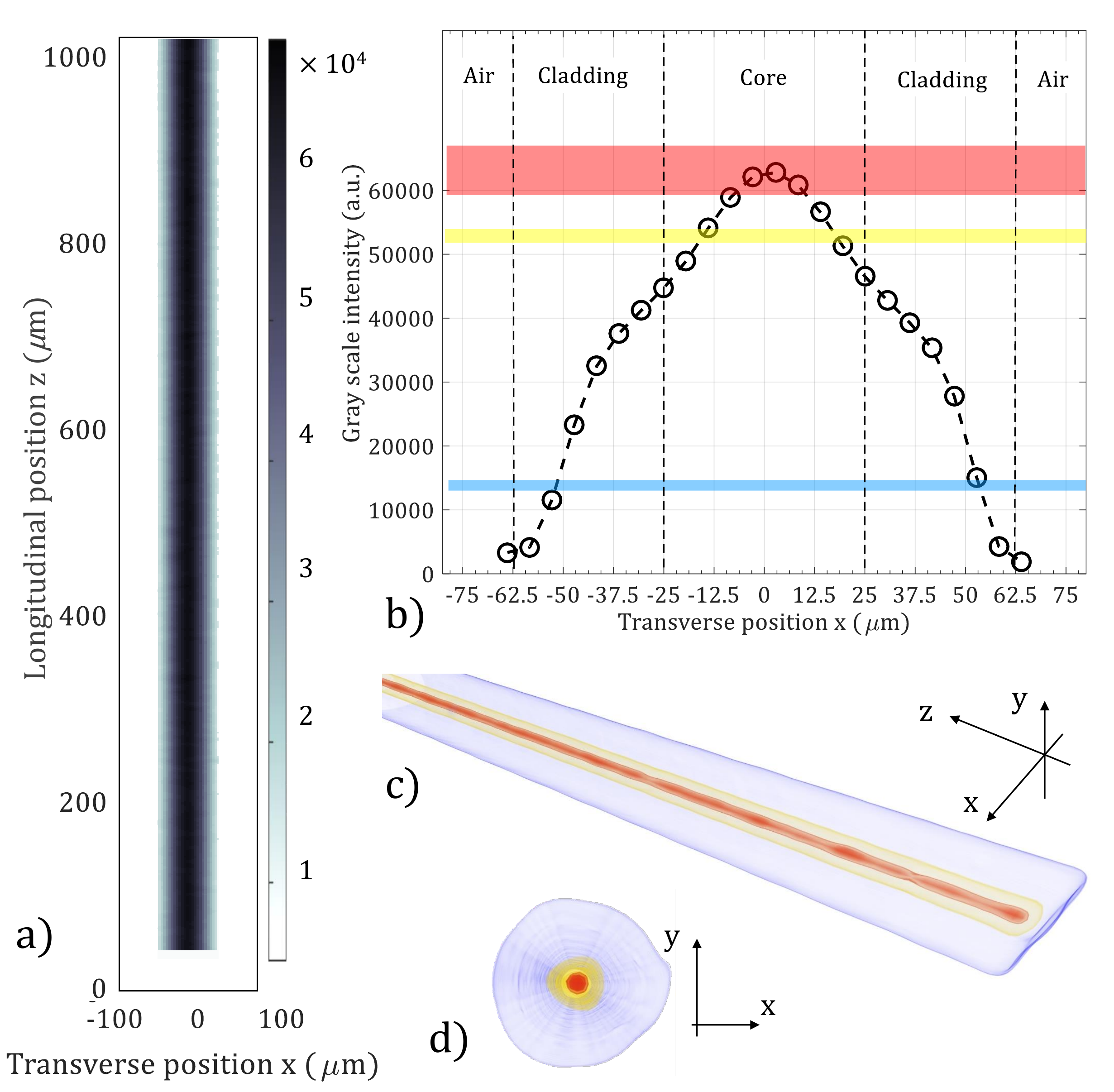}
\caption{X-ray imaging. (a) Radiography of a brand new GRIN 50/125 fiber. (b) Corresponding intensity profile along the transverse direction $x$, averaged over 200 $\mu$m in $z$. 
(c) 3D rendering of the $\mu$-CT image of a fiber, damaged by the exposition for 1.5 h to a laser pulse train at 1.03 $\mu$m wavelength and with 5.54 MW of peak power. Red, yellow and blue zones are obtained by filtering the $\mu$-CT intensity, as illustrated in panel (b). (d) Section of the fiber in the $x-y$ plane.}
\label{radiography}
\end{figure}

In Fig.\ref{radiography}a, we report a radiographic image of the tip of the fiber before any damaging has occurred. This image was obtained by averaging over 100 acquisitions, corresponding to a total exposure time of 1000 s. We used a microfocus source (Hamamatsu L12161-07), which emits a conical polychromatic, 10 W average power X-ray beam. The maximum energy of the X-photons was of 60 keV
. The intensity profile across the transverse dimension of the fiber, measured by a flat panel detector (Hamamatsu C7942SK-05) after averaging over a longitudinal path $z$ of 200 $\mu$m, is shown in Fig.\ref{radiography}b. 
From this figure we can appreciate that X-ray absorption is capable of distinguishing the core of the fiber from its cladding. Indeed, a change of convexity of the dashed curve, which interpolates the experimental data (void circles), suddenly appears when passing from the core to the cladding. 
Owing to the polychromatic nature of our X-ray source, we cannot determine the value of $g$ from the shape of the intensity profile, thus we are limited to a qualitative analysis. 


Nevertheless, results in Fig.\ref{radiography}a and b clearly demonstrate that the presence of index grading of the fiber core is captured by X-ray absorption measurements. This indicates that $\mu$-CT can be an appropriate tool for detecting the presence of fiber refractive index modifications. As a matter of fact, the absorption contrast $\mu$-CT is more sensitive to detecting step-like index variations, rather than gradual index changes. For this reason, we carried out $\mu$-CT imaging of a GRIN fiber, after its exposition to a laser pulse train (with peak power $P = 5.54$ MW at 1.03 $\mu$m) for a period of 1.5 hours. The 3D rendering of the $\mu$-CT image reconstruction is shown in Fig.\ref{radiography}c. The latter was obtained by means of the software Fiji \cite{schindelin2012fiji}. The $\mu$-CT intensity has been filtered, in order to highlight the presence of damages. We applied three different filters to the gray scale of the raw $\mu$-CT rendering. Each color corresponds to different intensity intervals, which are shown by horizontal bars in Fig.\ref{radiography}b. The red zone is enclosed inside the fiber core, and it includes the fiber axis. Whereas, the yellow zone, which is still fully located inside the core, does not contain the fiber axis. Finally, the blue zone involves the most peripheral part of the cladding. In order to better illustrate the three zones, a cross-section of the fiber in the $xy$ plane is shown in Fig.\ref{radiography}d.

\begin{figure}[ht!]
\centering\includegraphics[width=8.6cm]{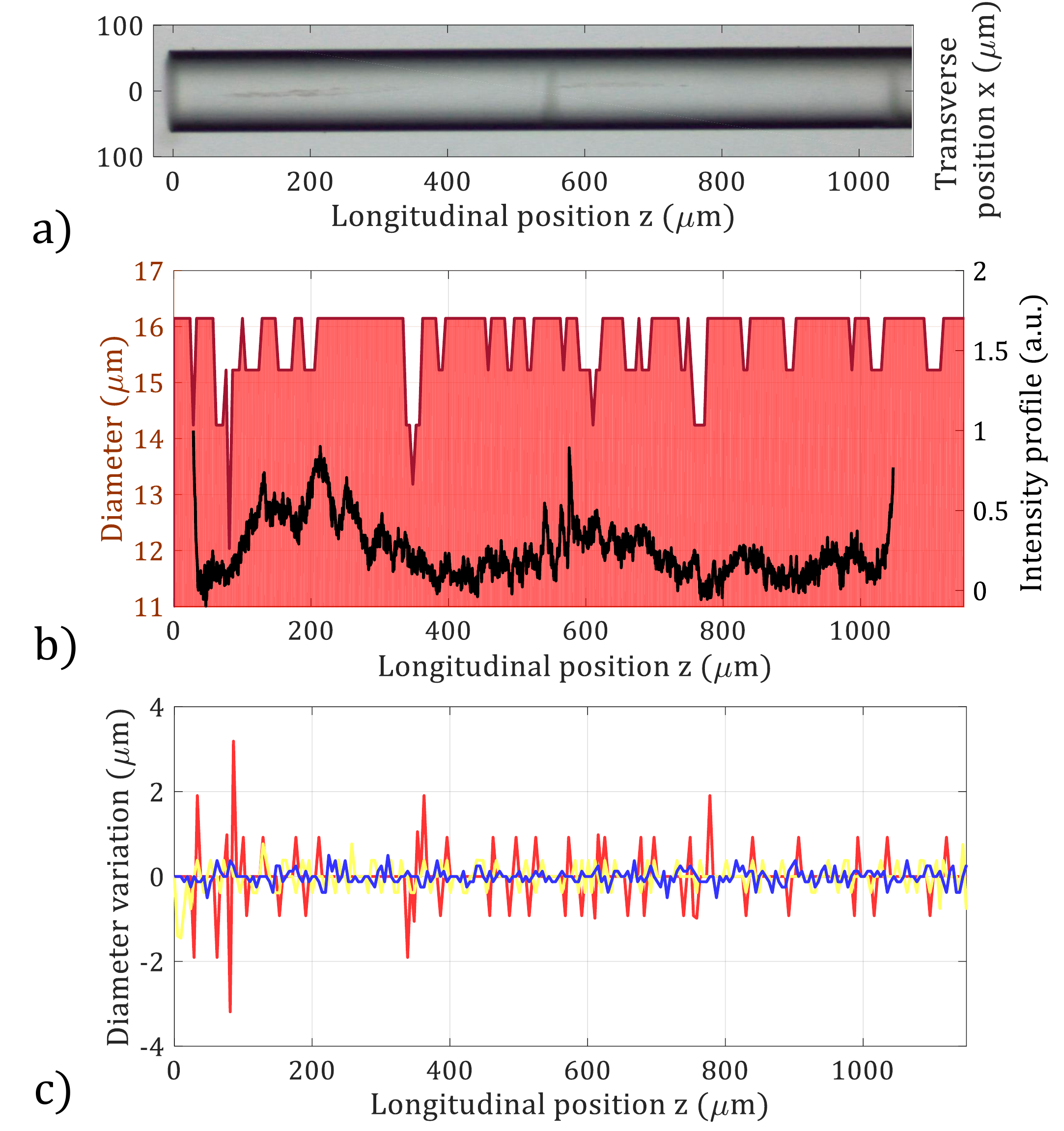}
\caption{Analysis of fiber damages. (a) Optical microscope image. (b) Comparison between the intensity profile along the fiber axis in (a) (black curve) with the diameter of the red zone (red area). (c) Variation along $z$ of the diameter of each zone.}
\label{tomography}
\end{figure}

Some interesting features, indicating the presence of damages, could indeed be detected. In particular, in Fig.\ref{radiography}c we may appreciate that, differently from the other two zones, the red zone experiences a local shrinkage. The relationship between the latter and the damages imaged by optical microscopy can be revealed by examining Fig.\ref{tomography}. Here, we show first the microscope image of the damaged fiber (Fig.\ref{tomography}a). 
Next, in Fig.\ref{tomography}b, we compare the intensity profile of the latter along the fiber axis (black curve), with the diameter of the $\mu$-CT red zone (red area). As it can be seen, the presence of a damage, which appears as a hill for the black curve,  corresponds to a steep reduction of the red zone diameter. Specifically, since $\mu$-CT is sensitive to refractive index mismatches, minima of the red curve correspond to edges of the damages, i.e., they are only visible whenever the black curve has a sufficiently abrupt variation. 
Finally, we highlight that the occurrence of diameter variations uniquely belongs to the red zone. As mentioned above, we found that the yellow and the blue zones have a nearly constant diameter, in the spite of some weak short-scale fluctuation (see Fig.\ref{tomography}c). Being far from the fiber axis, these zones are largely unaffected by the presence of damages.

In our measurements, we have approached the magnification limit of our $\mu$-CT system, which is only capable of detecting refractive index steps with about 7 $\mu$m of spatial resolution. Furthermore, given the spatial incoherence of our polychromatic source, we could not directly access information about variations of the real part of the linear fiber susceptibility, or refractive index.
Nevertheless, our results provide a proof of concept that X-ray absorption and scattering can be used for detecting micron-size damages in optical fibers, thus paving the way for their future investigation by means of phase-contrast measurements.

In conclusion, we studied the damages to standard optical fibers that are induced by MPI of intense femtosecond laser pulses, that propagate in either in normal or in the anomalous dispersion regime. Remarkably, by monitoring the output spectrum and the power transmission of the fiber, we observed that MPI damaging has strong macroscopic effects. Specifically, a slow temporal reshaping of SCG. Both the transmitted power and the SC spectra experience an irreversible temporal evolution towards equilibrium values. These are reached within a time scale of the order of a few hours, during which local modifications of the refractive index are progressively generated. The latter could be detected by means of optical microscopy and $\mu$-CT. Our results demonstrate that X-ray imaging, either based on absorption or scattering, can be a powerful, and previously unforeseen, tool for detecting micron-size features of glassy structures, such as standard optical fibers. This paves the way for future investigations, performed by means of coherent X-ray sources.

\section*{Funding}
European Research Council (740355); Ministero dell’Istruzione, dell’Università e della Ricerca (R18SPB8227, PIR01-00008); Ministry of Education and Science of the Russian Federation (14.Y26.31.0017); Russian Science Foundation (21-72-30024); Agence Nationale de la Recherche (ANR-18-CE080016-01, ANR-10-LABX-0074-01).
\section*{Acknowledgements}
We acknowledge the support of CILAS Company (ArianeGroup, X-LAS laboratory) and “Région Nouvelle Aquitaine” (F2MH and Nematum).

\section*{Disclosures}
The authors declare no conflict of interest.
\section*{Data availability}
Data underlying the results presented in this Letter are not publicly available at this time but may be obtained from the authors upon reasonable request.


\bibliography{biblio}






\end{document}